# Nuclear size corrections to the energy levels of single-electron atoms

Babak Nadiri Niri [a]

Research Institute for Astronomy and Astrophysics of Maragha (RIAAM), IRAN,
P. O. Box: 55134 - 441.

**Abstract**

A study is made of nuclear size corrections to the energy levels of single-electron atoms for the ground state of hydrogen like atoms. We consider Fermi charge distribution to the nucleus and calculate atomic energy level shift due to the finite size of the nucleus in the perturbation theory context. The exact relativistic correction based upon the available analytical calculations is compared to the result of first-order relativistic perturbation theory and the non-relativistic approximation. We find small discrepancies between our perturbative results and those obtained from exact relativistic calculation even for large nuclear charge number $Z$.

## 1. Introduction

As we know, the unphysical infinity in the $1/r$ potential at the origin makes it necessary that this potential be modified for values of $r$ inside a region about the origin that can be identified with the nucleus of the atom. The remedy is attributing finite size to the nucleus of the atom. The resulting correction due to the finite size of the nucleus leads to the shift of atomic energy levels. From another point of view, there is isotope shift of atomic energy levels due to this kind of corrections.

The dependence of the correction to the atomic energy level on the form of the potential energy inside the nucleus necessitates a choice of a model for the nuclear potential. For two common models, for nuclear potential function, which respectively simulate either a uniform charge distribution or a constant potential inside nucleus, the atomic energy level shift has been calculated [1].

[a] Author to whom any correspondence should be addressed
E-mail: bnbnadiri9@gmail.com
Tel: +98 9305645495   Fax :   +98 41 37412224

Calculation of these type of corrections have attracted a lot of attention. For a review see Ref [1-7]. The exact treatment of the problem is based on a solution of the Dirac equation for all values of $r$. The method reduces the computation of the energies of the electron, in interaction with a finite size nucleus, to a boundary value problem involving a single unknown eigenvalue [1]. In the present paper, we adopt another two appropriate charge distribution to the nucleus: Fermi and $1/r$ distributions; and calculate the correction for the ground state of electronic hydrogen like atom due to these charge distributions of nucleus (nuclear size). The main focus is on the comparison of the exact results with the results of two approximate methods. The approximate methods are perturbation theory and non-relativistic treatment as described in Section 3.

In Section 2, we briefly discuss the exact solution of Dirac equation in the presence of external potential. The approximate methods are described in Section 3.

In Section 4, the numerical calculation in perturbation theory is discussed. Finally our numerical results are compared with the results obtained from perturbation theory using both relativistic and non-relativistic wave functions for two physical charge distribution models to the nucleus.

## 2. EXACT CALCULATION

The solution of the Dirac equation in the presence of external potential leads to the coupled differential equations for the radial wave functions as [8]

$$\left[\frac{d}{dr}+\frac{1+\kappa}{r}\right]g(\mathrm{r})-\frac{1}{\hbar c}\left[E+mc^2-V(\mathrm{r})\right]f(\mathrm{r})=0 \qquad (1)$$

$$\left[\frac{d}{dr}+\frac{1-\kappa}{r}\right]f(\mathrm{r})-\frac{1}{\hbar c}\left[E-mc^2-V(\mathrm{r})\right]g(\mathrm{r})=0 \qquad (2)$$

Where $f(r)$ and $g(r)$ are the upper and lower components of the radial eigenfunctions, respectively; $E$ is the energy eigenvalue and κ is the eigenvalue of the operator $\hat{\kappa} = \hat{\sigma} \cdot \hat{L} + \hbar$. Here, for a given value of $j$, the quantum number κ has the possible values $\pm(j+\frac{1}{2})$ corresponding to values of $l$ and $l'$ equal to $j \pm \frac{1}{2}$ and $j \mp \frac{1}{2}$ respectively.

For values of radial coordinate $r$ greater than or equal to a value R which defines the nuclear radius, we assume that the central potential has the coulomb form,

$$V(r) = -\frac{Ze^2}{r}, \quad (r \geq R) \tag{3}$$

Solution of the radial Dirac equation for this region leads to the familiar formulae for the allowed energy eigenvalues of the electron given by

$$E = mc^2 \frac{\left(n' + \sqrt{\left(j+\frac{1}{2}\right)^2 - (aZ)^2}\right)}{\sqrt{\left(n' + \sqrt{\left(j+\frac{1}{2}\right)^2 - (aZ)^2}\right)^2 + (aZ)^2}}, \quad n' = 0,1,2,... \tag{4}$$

Where $\alpha = e^2/\hbar c \approx \frac{1}{137}$ is the fine-structure constant and $n' = n - j - \frac{1}{2}$ is defined from principal quantum number $n = 1, 2, \ldots$. It can be shown that the functions $g(r)$ and $f(r)$ have the explicit forms[1]:

$$g(r) = \frac{1}{2} N_\beta e^{-\rho/2} \rho^{\gamma-1} \Big[ {}_1F_1(\gamma - \zeta(E'), 2\gamma+1, \rho) - \eta_-(E') \, {}_1F_1(\gamma+1-\zeta(E'), 2\gamma+1, \rho)$$
$$-\Delta(\beta,\gamma) \rho^{-2\gamma} \, {}_1F_1(-\gamma - \zeta(E'), -2\gamma+1, \rho) - \Delta(\beta,\gamma) \eta_+(E') \rho^{-2\gamma} \, {}_1F_1(-\gamma+1-\zeta(E'), -2\gamma+1, \rho) \Big] \tag{5}$$

$$f(r) = \frac{1}{2} N_\beta \sqrt{\frac{1-E'}{1+E'}} e^{-\rho/2} \rho^{\gamma-1} \Big[ {}_1F_1(\gamma - \zeta(E'), 2\gamma+1, \rho) + \eta_-(E') \, {}_1F_1(\gamma+1-\zeta(E'), 2\gamma+1, \rho)$$
$$-\Delta(\beta,\gamma) \rho^{-2\gamma} \, {}_1F_1(-\gamma - \zeta(E'), -2\gamma+1, \rho) + \Delta(\beta,\gamma) \eta_+(E') \rho^{-2\gamma} \, {}_1F_1(-\gamma+1-\zeta(E'), -2\gamma+1, \rho) \Big] \tag{6}$$

Where $N_\beta$ represents a normalization constant and the energy parameter $E'$ must be derived from the continuity conditions at $r = R$. Beside, we have used the following notation

$$E' \equiv \frac{E}{mc^2} \quad , \quad \zeta(E') \equiv \frac{\alpha Z E'}{\sqrt{1-E'^2}} \quad , \quad \eta_{\pm}(E') \equiv \frac{\gamma \pm \zeta(E')}{\kappa - \zeta(E')/E'} \tag{7}$$

$$\gamma^2 = \left(j+\frac{1}{2}\right)^2 - (\alpha Z)^2 \quad \text{and} \quad \beta = \beta_{\pm} = \alpha Z \frac{E}{\sqrt{(mc^2)^2 - E^2}} \pm \frac{1}{2}$$

For values of the radial coordinate less than the nuclear radius, the radial Dirac equations have been calculated analytically for two common models in [1], such as 1) uniformly charged nucleus and 2) constant potential inside nucleus.

The solutions of the Dirac equation for values of $r$ exterior and interior to the nucleus need to be made continuous at the boundary of the nucleus defined by $r = R$. The continuity requirement at $r = R$ produces the simultaneous equations:

$$g_{\text{interior}}(R) = g_{\text{exterior}}(R), \quad f_{\text{interior}}(R) = f_{\text{exterior}}(R) \tag{8}$$

which can be conveniently combined into the "matching equation" as

$$\frac{g_{\text{interior}}(R)}{f_{\text{interior}}(R)} = \frac{g_{\text{exterior}}(R)}{f_{\text{exterior}}(R)} \tag{9}$$

The equation has the effect of reducing the computation of the energies of the atomic electron in the case of a finite size nucleus to a boundary value problem involving a single unknown $E'$, the solution of which determines the allowed energy eigenvalues.

## 3. APPROXIMATE METHODS

We can compare the energy eigenvalues derived from equation (9) with the corrected eigenvalues obtained from the first-order perturbation theory under the assumption that the change in the coulomb potential in the interior of the nucleus is treated as a perturbation to the Hamiltonian as, $H = H_0 + V + \Delta V(r)$; in which $V = -Ze^2/r$ and

$$\Delta V(r) = \begin{cases} 0 & , \quad r > R \\ \Delta(r) & , \quad r \leq R \end{cases} \tag{10}$$

For spherically symmetric charge distribution (model 1) inside the nucleus, the relation

$$V(r) = \int_{-\infty}^{R} E_{out}\, dr + \int_{R}^{r} E_{in}\, dr \text{ yields}$$

$$\Delta(r) = \frac{Ze^2}{R}\left(\frac{r^2}{2R^2} - \frac{3}{2} + \frac{R}{r}\right), \tag{11}$$

In which we have substituted the well-known formulas for electric fields inside and outside the sphere:

$$E(r) = \begin{cases} \dfrac{\rho R^3}{3\varepsilon_0 r^2}, & r > R \\ \dfrac{\rho}{3\varepsilon_0} r, & r \leq R \end{cases} \tag{12}$$

Performing similar straightforward calculations for the constant potential inside the nucleus (model 2), one gets

$$\Delta(r) = -\frac{Ze^2}{r} + \frac{Ze^2}{R}, \tag{13}$$

Now we want to obtain the energy shift of the ground state ($n' = 0$) atomic electron, in which we have assumed uniform charge distribution inside the nucleus. From first order perturbation theory $\Delta E = \langle \psi^0 | \Delta V(r) | \psi^0 \rangle$, and using relativistic (Dirac) form for $\psi^0(\vec{r})$,

$$\psi^{(0)}(\vec{r}) = \begin{pmatrix} ig(r)\Omega_{jlm_j} \\ f(r)\Omega_{jl'm_j} \end{pmatrix} \tag{14}$$

where $\Omega_{jlm_j}$ is the Dirac spinor, we obtain

$$\Delta E = \int_0^\infty r^2 dr \int_{4\pi} d\Omega_{\hat{r}}\, \overline{\psi}^{(0)}(\vec{r}) \Delta V(r) \psi^{(0)}(\vec{r})$$
$$= \int_0^\infty r^2 dr \left[g^2(r) + f^2(r)\right] \Delta V(r) \underbrace{\int_{4\pi} d\Omega_{\hat{r}}\, \overline{\Omega}_{jlm_j} \Omega_{jlm_j}}_{=1} \tag{15}$$

On the other hand, $f(r)$ and $g(r)$ is derived from relations (5) and (6) for ground state of atomic electron as

$$g(r) = N\, r^{\gamma-1} e^{-qr}(-\kappa + 1) \quad , \quad f(r) = -N\, \frac{\alpha Z/2}{\sqrt{1-\left(\frac{\alpha Z}{2}\right)^2}} r^{\gamma-1} e^{-qr}(-\kappa + 1) \tag{16}$$

Where

$$q = \frac{\sqrt{(mc^2)^2 - E^2}}{\hbar c} \quad , \quad \gamma = \sqrt{\kappa^2 - (\alpha Z)^2} \quad , \quad \kappa_{ground\ state} = -1 \tag{17}$$

Using ground-state energy eigenvalue $E$,

$$E = \frac{mc^2}{\sqrt{1 + \frac{(\alpha Z)^2}{\gamma^2}}} \quad , \tag{18}$$

$q$, may be written as

$$q = \frac{\alpha Z}{\sqrt{\gamma^2 + (\alpha Z)^2}} \frac{mc^2}{\hbar c} = \frac{\alpha Z}{|\kappa|} \frac{mc}{\hbar} = \frac{\alpha Z}{\lambda_c} \tag{19}$$

Where $\lambda_c$ is the Compton wavelength of the electron. From relations (11) and (15), $\Delta E$ takes the form

$$\Delta E = \frac{Ze^2}{R} \int_0^R r^2 dr \left[ g^2(r) + f^2(r) \right] \left( \frac{r^2}{2R^2} - \frac{3}{2} + \frac{R}{r} \right)$$

$$= 4N^2 \frac{Ze^2}{R} \left[ 1 + \frac{\alpha Z/2}{1 - \left(\frac{\alpha Z}{2}\right)^2} \right] \int_0^R dr \left( \frac{r^{2\gamma+2}}{2R^2} - \frac{3}{2} r^{2\gamma} + R r^{2\gamma-1} \right) e^{-2qr} \quad , \tag{20}$$

Beside, for value of $r \leq R$ we have $qr \leq \frac{\alpha Z}{\lambda_c} R = \alpha Z R' \ll 1$; So one can consider $e^{-2qr} \approx 1$. Therefore

$$\Delta E = \frac{4N^2}{\left[1 - \left(\frac{\alpha Z}{2}\right)^2\right]} \frac{Ze^2}{R} \int_0^R dr \left( \frac{r^{2\gamma+2}}{2R^2} - \frac{3}{2} r^{2\gamma} + R r^{2\gamma-1} \right) \quad , \tag{21}$$

$$= \frac{2N^2 Ze^2}{\left[1 - \left(\frac{\alpha Z}{2}\right)^2\right]} R^{2\gamma} \left( \frac{1}{2\gamma+3} - \frac{3}{2\gamma+1} + \frac{1}{\gamma} \right)$$

Here the normalization constant can be obtained according to the prescription $\int \psi^\dagger \psi \, dV = 1$. So one gets from relations (14), (16) and (19)

$$\frac{4N^2}{\left[1 - \left(\frac{\alpha Z}{2}\right)^2\right]} \int_0^\infty dr \, r^{2\gamma} e^{-2\frac{\alpha Z}{\lambda_c} r} = \frac{4N^2}{\left[1 - \left(\frac{\alpha Z}{2}\right)^2\right]} \frac{\Gamma(2\gamma+1)}{\left(\frac{2\alpha Z}{\lambda_c}\right)^{2\gamma+1}} = 1 \tag{22}$$

Then

$$N^2 = \frac{\left[1 - \left(\frac{\alpha Z}{2}\right)^2\right]}{4\Gamma(2\gamma+1)} \left(\frac{2\alpha Z}{\lambda_c}\right)^{2\gamma+1} \tag{23}$$

Substituting the expression (23) in relation (21), we obtain

$$\Delta E = \frac{Ze^2}{2} \frac{\left(\frac{2\alpha Z}{\lambda_c}\right)^{2\gamma+1}}{\Gamma(2\gamma+1)} R^{2\gamma} \left[\frac{1}{2\gamma+3} - \frac{3}{2\gamma+1} + \frac{1}{\gamma}\right] \qquad (24)$$

Now, the following relations

$$Ze^2 = \frac{Ze^2}{\hbar c} \frac{mc^2}{mc/\hbar} = \alpha Z\, mc^2\, \lambda_c \quad , \quad R = R'\lambda_c \quad , \qquad (25)$$

Allow us to write (24) as [1]:

$$\Delta E = \frac{(\alpha Z)^2 (2\alpha Z)^{2\gamma}}{\Gamma(2\gamma+1)} R'^{2\gamma}\, mc^2 \left(\frac{1}{2\gamma+3} - \frac{3}{2\gamma+1} + \frac{1}{\gamma}\right) \qquad (26)$$

In comparison, the non-relativistic calculation gives the result [9]:

$$\Delta E = \frac{2}{5}(\alpha Z)^4 R'^2 mc^2 \qquad (27)$$

For small values of $Z$, in the approximation in which $\gamma = \sqrt{1-(\alpha Z)^2}$ reduces to one, the relativistic result (26) coincides with the non-relativistic result (27).

Using the empirical relation

$$R \cong r_0 A^{1/3} \quad , \quad r_0 = 1.2 \times 10^{-15} m \qquad (28)$$

Where $r_0$ and $A$ are the nucleus radius and mass number respectively, the relation (26) recasts in the form

$$\Delta E = \frac{(6.2150517 \times 10^{-3})^{2\gamma}}{\Gamma(2\gamma+1)} \left(\frac{1}{2\gamma+3} - \frac{3}{2\gamma+1} + \frac{1}{\gamma}\right) mc^2 (\alpha Z)^{2+2\gamma} \cdot A^{2\gamma/3} \qquad (29)$$

For constant potential inside nucleus the approach is similar. In the present work, for comparison, we adopt another two models for nuclear charge distribution: $1/r$ charge distribution and a Fermi charge distribution for $r < R$. With $\Delta(r)$ defined by either of the two formula

$$\Delta(r) = \frac{Ze^2}{R}\left(\frac{R}{r} + \frac{r}{R} - 2\right), \quad 1/r - ch\text{arge distribution} \qquad (30)$$

$$\Delta(r) = -e\int_R^\infty \frac{4\pi\rho_0}{r^2}\left[\int_0^R \frac{r^2 dr}{1+\exp\left(\frac{r-c}{k}\right)}\right] dr + e\int_R^r \frac{4\pi\rho_0}{r^2}\left[\int_0^r \frac{r^2 dr}{1+\exp\left(\frac{r-c}{k}\right)}\right] dr + \frac{Ze^2}{r}, \quad \textit{Fermi ch}\text{arge distribution} \tag{31}$$

And using the relation

$$\Delta E = \int_0^\infty r^2 \left[g^2(r) + f^2(r)\right] \Delta(r)\, dr \tag{32}$$

Along with the form for $\Delta(r)$ in equations (30-31), results in corrections to the energy of ground state given by the respective formulas

$$\Delta E = \frac{(\alpha Z)^2 (2\alpha Z)^{2\gamma}}{\Gamma(2\gamma+1)} R'^{2\gamma} mc^2 \left(-\frac{2}{2\gamma+1} + \frac{1}{2\gamma+2} + \frac{1}{2\gamma}\right), \quad \frac{1}{r} - \textit{charge distribution}. \tag{33}$$

$$\Delta E = \frac{\left(\frac{2\alpha Z}{\lambda_c}\right)^{2\gamma+1}}{\Gamma(2\gamma+1)} \int_0^R dr\, r^{2\gamma}\, \Delta V(r), \quad \textit{Fermi ch}\text{arge distribution}. \tag{34}$$

Clearly, relation (31) is being substituted for $\Delta V(r)$ in the expression (34). Here, we consider

$$\rho_{\frac{1}{r}}(r) = \frac{\rho_0}{r}, \qquad \rho_{Fermi}(r) = \frac{\rho_0}{1+\exp\left(\frac{r-c}{k}\right)} \tag{35}$$

The two parameters $c$ and $k$ are determined, for instance, by fitting to densities derived from measured form factors [10-11]; and the factor $\rho_0$ is given by normalization condition

$$\int_0^R \rho(r)\, dr = Ze \tag{36}$$

## 4. CONCLUSIONS

The dependence of the correction to the energy on the form of the potential energy inside the nucleus necessitates a choice of a model for the nuclear potential. For two common models, for nuclear potential function, which respectively simulate either a uniform charge distribution or a constant potential inside nucleus, $\Delta E_{exact}$ and $\Delta E_{Pert.}$ have been calculated [1].

| Z | A | $\Delta E_{exact}$ (eV) | $\Delta E_{Perturbation}$ (eV) (constant potential) | $\Delta E_{Perturbation}$ (eV) (uniform charge dist.) | $\Delta E_{code}$ (eV) | $\Delta E_{non-relativistic}$ (eV) |
|---|---|---|---|---|---|---|
| 1 | 1 | $5.60 \times 10^{-9}$ | $9.33 \times 10^{-9}$ | $5.60 \times 10^{-9}$ | $5.60 \times 10^{-9}$ | $5.60 \times 10^{-9}$ |
| 1 | 2 | $8.89 \times 10^{-9}$ | $14.82 \times 10^{-9}$ | $8.89 \times 10^{-9}$ | $8.89 \times 10^{-9}$ | $14.81 \times 10^{-9}$ |
| | | | | | | |
| 47 | 107 | 1.26755 | 2.21598 | 1.36265 | 1.26706 | 0.61558 |
| 47 | 109 | 1.28288 | 2.24182 | 1.37854 | 1.28179 | 0.62323 |
| | | | | | | |
| 63 | 151 | 8.60942 | 15.75194 | 9.89420 | 8.60438 | 2.50026 |
| 63 | 153 | 8.67636 | 15.87512 | 9.97157 | 8.67124 | 2.52228 |
| | | | | | | |
| 81 | 203 | 60.383 | 120.540 | 78.388 | 60.342 | 8.322 |
| 81 | 205 | 60.698 | 121.177 | 78.802 | 60.658 | 8.377 |
| | | | | | | |
| 92 | 235 | 193.180 | 420.373 | 281.357 | 193.066 | 15.270 |
| 92 | 238 | 194.376 | 423.016 | 283.126 | 194.260 | 15.400 |

Table 1: Values derived from the present calculation and from relativistic and non-relativistic perturbation theory for correction to the ground state energy of an hydrogenic atom produced by the finite size of a nucleus of charge Z. Models 1 and 2 assume (1) a uniformly charged nucleus and (2) a constant potential inside the nucleus, respectively [1]. Ground state : $n' = 0$, $k = -1$.

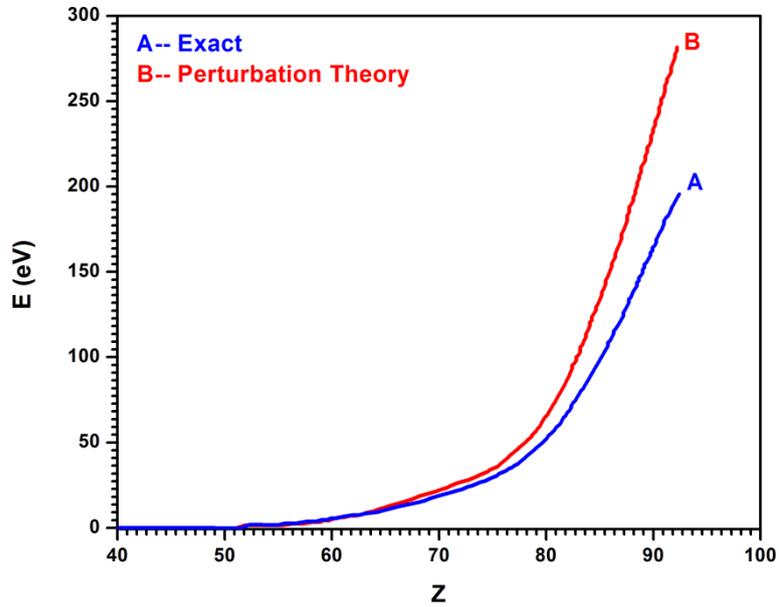

Figure 1: Graphs versus Z of the nuclear size correction to the ground state energy of a hydrogenic atom obtained from matching condition in equation (8) (Curve A), and relativistic perturbation theory (Curve B) using model 1 [1]

For these two models, Table1 lists calculated values of the energy correction to the ground states of single-electron atoms corresponding to stable isotopes of the five elements $H$, $U$, $Ag$, $Eu$ and $Tl$. In particular, the table compares the corrections, $\Delta E$, derived from the matching condition in equation (9) with the values of $\Delta E$ obtained for the same model of the nuclear potential, using first-order perturbation theory based on both relativistic $\Delta E_{pert.}$ and non-relativistic wave functions $\Delta E_{non-rel.}$. The different values for $\Delta E$ as a function of $Z$ predicted by the perturbation theory and exact theory, for these two models, are summarized by the graphs in figure 1.

It is useful to compare the results derived from the matching condition in equation (9) with the results extracted from an atomic structure code for the same model. To do this, we include in table 1 the values of $\Delta E$ obtained from general purpose relativistic atomic structure program, GRASP [12], for the case of a uniformly charged nucleus. Comparison of these values, denoted by $\Delta E_{code}$ with the values obtained from equation (9) shows that the two sets of values are in excellent agreement.

In analogy with the results listed in table (1), we list in table (2) the calculated values of $\Delta E$ for the ground states of electronic atoms with Fermi and $1/r$ charge distribution. As expected, results of these two models are in excellent agreement with $\Delta E_{code}$ and $\Delta E_{pert.}$. In comparison with previous models, we find better results for $\Delta E_{pert.}$.

The different values for $\Delta E$ as a function of $Z$ predicted by the perturbation theory and exact theory (for $1/r$ and Fermi charge distribution) are summarized by the graphs in figure (2). In spite of that the $1/r$ charge distribution is not of much physical interest, the related results is in good agreement with $\Delta E_{code}$ and $\Delta E_{exact}$.

In summary, for values of Z greater than 40 in the case of electronic atoms, we find large discrepancies between our results and those obtained from first-order perturbation theory using relativistic wave functions. But with considering physical models (Fermi charge distribution) to the nucleus we find small discrepancies between perturbative and exact results even for large nuclear charge number $Z$.

| Z | A | $\Delta E_{exact}$ (eV) (Uniform) | $\Delta E_{Perturbation}$ (eV) (Uniform charge dist.) | $\Delta E_{Perturbation}$ (eV) $1/r$-charge dist. | $\Delta E_{Perturbation}$ (eV) (Fermi-charge dist.) | $\Delta E_{code}$ (eV) |
|---|---|---|---|---|---|---|
| 1 | 1 | $5.60 \times 10^{-9}$ | $5.60 \times 10^{-9}$ | $4.60 \times 10^{-9}$ | $5.60 \times 10^{-9}$ | $5.60 \times 10^{-9}$ |
| 1 | 2 | $8.89 \times 10^{-9}$ | $8.89 \times 10^{-9}$ | $7.41 \times 10^{-9}$ | $8.89 \times 10^{-9}$ | $8.89 \times 10^{-9}$ |
| | | | | | | |
| 47 | 107 | 1.26755 | 1.36265 | 1.1443173 | 1.351 | 1.26276 |
| 47 | 109 | 1.28288 | 1.37854 | 1.15761 | 1.362 | 1.28179 |
| | | | | | | |
| 63 | 151 | 8.60942 | 9.89420 | 8.357759 | 8.923 | 8.60438 |
| 63 | 153 | 8.67636 | 9.97157 | 8.423116 | 9.021 | 8.67124 |
| | | | | | | |
| 81 | 203 | 60.383 | 78.388 | 47.771 | 71.62 | 60.342 |
| 81 | 205 | 60.698 | 78.802 | 48.203 | 72.78 | 60.658 |
| | | | | | | |
| 92 | 235 | 193.180 | 281.357 | 25.3 | 242.51 | 193.066 |
| 92 | 238 | 194.376 | 283.126 | 253.622 | 244.12 | 194.260 |

Table 2: Values derived from the present calculation and from relativistic and non-relativistic perturbation theory for correction to the ground state energy of a hydrogenic atom produced by the finite size of a nucleus of charge Z. Assuming (1) $\frac{1}{r}$ charged nucleus and (2) a Fermi charge distribution inside the nucleus, respectively. Ground state : $n' = 0$ , $k = -1$.

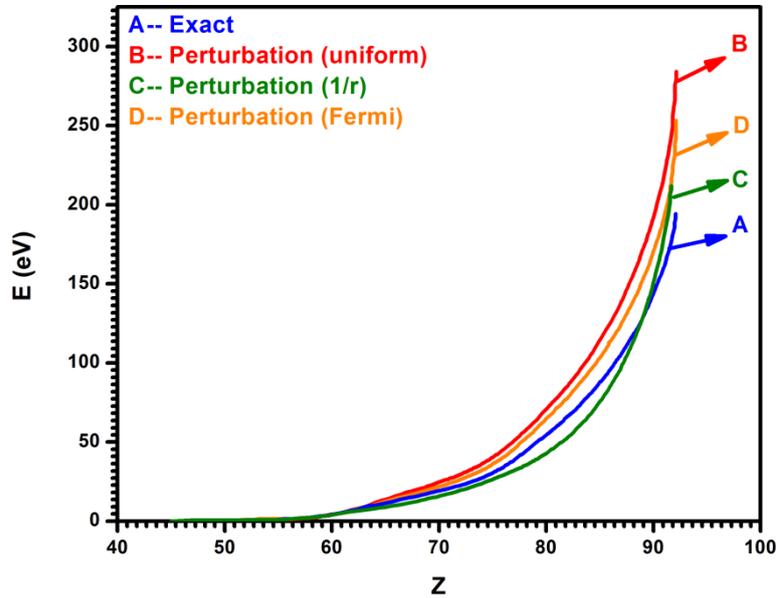

Figure 2: Graphs versus Z of the nuclear size correction to the ground state energy of a hydrogenic atom obtained from matching condition in equation (9) (Curve A) and relativistic perturbation theory using model 1(Curve B), Curve C and Curve D for $\frac{1}{r}$ and Fermi charge distribution models respectively.

**Acknowledgments**

This work has been supported financially by the Research Institute for Astronomy and Astrophysics of Maragha (RIAAM).